\documentclass[12pt]{article}

 \usepackage[T1]{fontenc}
 \usepackage{lmodern}
 \usepackage[french]{babel}

\usepackage{geometry}
\usepackage{color}
\usepackage{srcltx}
\usepackage{epsfig}
\usepackage{graphicx}

\usepackage{amssymb}
\usepackage{amsfonts}
\usepackage{amsmath}
\usepackage{amsthm}
\usepackage{enumerate}
\usepackage{multicol}

\newtheorem{theorem}{Theorem}[section]
\newtheorem{proposition}[theorem]{Proposition}

\newcommand\cE{{\cal E}}

\newcommand\cF{{\cal F}}

\newcommand\cN{{\cal N}}

\newcommand\ve{\varepsilon}

\def\bbr{{\mathbb R}}

\def\text#1{\hbox{#1}}

\def\E{{\bf E}}
\def\P{{\bf P}}

\def\d{\mathrm{d}}
\def\build #1_#2{\mathrel{\mathop{\kern 0pt #1}\limits_{#2}}}
\newcommand{\zs}[1]{{\mathchoice{#1}{#1}{\lower.25ex\hbox{$\scriptstyle#1$}}
{\lower0.25ex\hbox{$\scriptscriptstyle#1$}}}}
\title{Evaluation d'une option asiatique dans le cadre des marchés de Lévy de type NIG et Variance gamma}
\author{Belkacem Berdjane\footnote{Département des mahtématiques, Université de Montréal, Québec, CANADA, berdjane\_b@yahoo.fr}}

\date{2017/06/05}

\begin{document}
\maketitle
%%%%%%%%%%%%%%%%%%%%%%%%%%%%%%%%%%%%%%%%%%%%%%%%%%%%%%%%%%%%%%%%%%%%%%%%%%%%%%%%%%%%%%%%%%%%%%%%%%%%
%%%%%%%%%%%%%%%%%%%%%%%%%%%%%%%%%%%%%%%%%%%%%%%%%%%%%%%%%%%%%%%%%%%%%%%%%%%%%%%%%%%%%%%%%%%%%%%%%%%%

\abstract{Dans ce travail on étudie la valeur d'une option asiatique dans le cas des modèles exponentiels de Lévy. Plus particulièrement on s'interesse au modèle NIG (normal inverse gaussien) et le modèle VG (variance gamma). Les modèles exponentiels de Lévy produisent des marchés incomplets. Ils existe donc une infinité de mesures  équivalentes sous lesquels les prix actualisés sont des martingales. On va s'intéresser à deux méthodes de construction d'une de ces mesures. La première est basée sur la transformée d'Esscher, et l'autre consiste à apporter une correction "risque-neutre" sur la dynamique des trajectoires. Ils s'avère, suivant les résultats numériques obtenus, que les deux méthodes produisent en général les même prix.  }

\vspace*{2mm}
{\sl Mots clé : } Modèles exponentiels de Lévy, Théorème fondamental, Produits dérivés, Transformée d'Esscher

\newpage

\section{Introduction}
Le modèle de marché standard de Black-Scholes est le modèle le plus populaire et le plus utilisé par les institutions financières. Dans ce modèle, le cours $S_t$ est un processus à trajectoire continues et à accroissements relatifs indépendants et stationnaires. Les rendement de $S_t$ suivent une loi log-normale \emph{i.e.} $\log (S_t/S_{t-1}) \rightsquigarrow \cN( (\mu- \sigma^2/2),\sigma^2 )$. Ces hypothèses "simplistes" conduisent à un marché complet qui donnent lieu à des formules fermées, par exemple pour le calcul de la valeur d'une option européenne.

\bigskip
En fait, le théorème fondamental de l'asset pricing" (voir \cite{DelbaenSchachermayer1994}) traduit la complétude du marché par l'unicité de la MME (mesure martingale équivalente) $P^*$, appelée également mesure risque-neutre, sous laquelle les prix actualisés sont une $\cF$-martingale. Cette mesure permet donc d'écrie la valeur d'une option de type européenne, de fonction de payement $f$ sous la forme :
\begin{equation}\label{form-call-eur}
  C_t=e^{-r \, (T-t)}\E^*\left( f(S_{u, t\le u \le T}) | \cF_t \right)
\end{equation}

\noindent où $E^*$ désigne l'espérance sous la mesure risque-neutre. Lorsque on à faire à une call européen $ f(S_{u, t\le u \le T})=(S_T-K)_+$, la résolution de cette équation peut se faire de façon explicite, et donne lieu à la célèbre formule de Black-Scholes.

\bigskip

Bien que le modèle de Black-Scholes soit très pratique et très utilisé, il se trouve qu'il ne répond pas à la réalité statistique observée e.g. \cite{EberleinKeller95}
et \cite{Rydberg200}.

Toutes les études numériques effectuées sur les données de marchés financiers \emph{invalident} le modèle log-normal. On peut citer par exemple l'asymétrie de la distribution empiriques de $\log (S_t)$ et le fait que les queues des distributions de $S_t$ sont plus épaisses que log-normal (e.g. \cite{Rydberg200}, \cite{Barndorff98}, \cite{MadanMilne91}).

\bigskip

Dans des travaux de recherches récents, beaucoup de chercheurs se sont intéressé à des alternatives au modèle standard de Black-Scholes. Des modèles plus représentatifs des données statistiques observées sur les marchés. On peut citer à titre d'exemples les modèles à volatilité stochastique et ceux basés sur les processus de Lévy (voir : Berdjane \& Pergamenschikov \cite{BePe-2013}\cite{BePe-2015}, Boyarchenko \& Levendorskii \cite{Boyarchenko02}, Chan \cite{Chan99} et Schoutens \cite{Schoutens03} etc.)

\bigskip

 Les modèles alternatifs les plus populaires sont probablement ceux basés sur les processus de Lévy. Une classe importante de ces processus est la classe GH (Hyperbolique généralisée) qui a été introduite pour la première fois en 1977 dans \cite{BarndorfNielsen77}. Les modèle NIG (Normal Inverse Gaussienne), et les modèles VG (Variance Gamma) font parties de la classe GH est sont investis par de nombreux chercheurs.  Le modèle NIG en particulier, est devenu un centre d'intérêt important à cause de sa flexibilité (voir Barndorff \& Nielsen \cite{Barndorff98}).

\bigskip
Les conséquences sur l'évaluation des options, de l'introduction de ces nouveaux modèles, sont multiples. La plus importante serait probablement la perte de la complétude du marché (et donc la perte de l'unicité de la mesure risque-neutre). Etant donné qu'on a pas unicité de la MME,
il est légitime de se poser la question de savoir, si deux différentes MME produisent les mêmes prix. Dans ce rapport on va tenter d'apporter une réponse à cette question, à travers les modèle NIG et VG, qu'on va utiliser pour l'évaluation d'une option asiatique.

\bigskip

Ce travail est organisé comme suit: La section 1 est l'introduction, la section \ref{sec-levy} est consacrée au processus de Lévy et à toute la théorie qui nous concerne. On commencera par rappeler les définitions de processus de Lévy, puis par l'introduction des modèles de marché de type exponentiel de Lévy. La section  \ref{sec-MME} et \ref{sec-transfE} seront consacrée à la construction de la MME à l'aide de la transformée d'Esscher. On présentera dans la section \ref{sec-NIG} la transformée d'Esscher pour un processus NIG, et dans la section \ref{sec-VG}, celle d'un processus de type VG. La section \ref{sec-simulation} sera consacrée au résulats numériques obtenus. La section \ref{sec-conclusion} fera l'objet de la conclusion.

\section{Modèles Exponentiels de Lévy}\label{sec-levy}

Les modèles exponentiels de Lévy constituent une généralisation du modèle classique de Black-Scholes en autorisant des sauts dans les trajectoires des prix. Ils s'avèrent que ces modèles sont une meilleure représentation de la réalité statistique observée sur les marchés, pour plusieurs raisons. Premièrement les prix peuvent sauter soudainement (événement particulier, crash,...) et certains risque de marché ne peuvent pas être considérés dans le cadre des modèles à trajectoires continues. Deuxièmement, le phénomène du "smile de volatilité", l'asymétrie des logarithme des rendement, ainsi que l'épaisseur des queues de leurs distributions, qui sont des propriétés observées dans les données de marché, peuvent bien être représentés par les modèles exponentiels de Lévy; ce qui n'est pas le cas du modèle classique de Black-Scholes (voir \cite{Tankov04}).
\bigskip

Un processus de Lévy de triplet $(a,\sigma,\nu)$ est un processus  à trajectoires discontinues (stochastiquement continu), et à accroissements indépendants et stationnaires. Ces conditions suffisent à assurer l'existence de l'exposant caractéristique $\kappa(\xi)$ dont la forme est donnée par la représentation de Lévy-Khintchine de sa fonction caractéristique :
$$
\E (e^{i \xi X_t})=e^{t \kappa(t \xi)}=\exp\left\{t \left( i a \xi -\frac{\sigma^2 \xi^2}{2}+\int_{\bbr} (e^{i \xi x}-1-i\xi x 1_{|x|<1})\nu(dx)\right)\right\}
$$
Les processus de Lévy sont essentiellement des processus avec sauts, car on peut démontrer (voir \cite{Tankov04}) que tout processus de Lévy continu est un mouvement brownien avec drift.

\bigskip

Pour assurer la positivité des prix, on modélise souvent les cours d'actions comme exponentielles de processus de Lévy
\begin{equation}\label{form-Levy-model}
   S_t =S_0 e^{rt+X_t}
 \end{equation}
Pour appliquer ce modèle à l'évaluation des options, il faudrait montrer l'existence d'une mesure martingale équivalente, afin d'avoir la viabilité du marché (Théorème fondamental de l'"asset pricing"). Lorsque le processus de Lévy $X_t$ est un MB ou un processus de Poisson, les prix d'actifs basés sur ce modèle donnent lieu à un marché complet (unicité de la mesure martingale). Dans tous les autres cas le marché est incomplet (voir \cite{Chan99} et \cite{Cherny01}). Mais avant de parler de viabilité de complétude du modèle, nous avons  besoins de savoir si deux probabilités sont équivalentes ou pas,  lorsqu'on connaît les caractéristiques de deux processus de Lévy différents.

\subsection{Mesure Martingale équivalente} \label{sec-MME}

Pour s'assurer qu'un modèle de Lévy est souhaitable pour une modélisation de marchés financiers, on a besoin de s'assurer qu'il ne permet pas d'opportunités d'arbitrage. On d'autre terme, il doit permettre l'existence d'une mesure martingale équivalente (MME) voir e.g. \cite{Selivanv05}, \cite{Jak02}, \cite{ChernyShiryaev02}.

\bigskip

Dans le cas du modèle de Black-Scholes, l'unicité de la mesure martingale peut être obtenue en changeant le drift du brownien. Dans les modèles de sauts purs, ceci n'est pas possible, mais une grande variété de mesures équivalentes peut être obtenue. La proposition suivante décrit le changement de mesure sous laquelle un processus de Lévy reste un processus de Lévy.

\begin{proposition}[voir \cite{Sato99} Théorèmes 33.1 et 33.2] Soit $(X,P)$ un processus de Lévy, à valeurs réelles, et soit $(a,\sigma,\nu)$ sont triplet caractéristique. Soit $\eta \in \bbr$ et $\phi :\bbr \to \bbr$ tel que
$$
\int_\bbr (e^{\phi(x)/2}-1)^2 \nu (dx) < \infty
$$
et soit
$$
U_t:=\eta X^c+\int_{0}^{t}\int_{\bbr} (e^{\phi(x)}-1)\tilde{J}_X(\d s \d x)
$$
où $X^c$ désigne la partie continue (Mouvement brownien) de $X$, et $\tilde{J}_X$ c'est la mesure de sauts compensée de X.
Le processus $\cE(U)_t$ (l'exponentielle stochastique de $U$) est une martingale positive telle que la mesure de probabilité $P'$ définie par
$$
\frac{\d \P'|_{\cF_t}}{\d \P|_{\cF_t}}=\cE(U)_t
$$
est équivalente à $\P$. Sous la mesure $P'$, le processus $X$ est un processus de Lévy de triplet caractéristique $(a,\sigma',\nu')$ où
\begin{eqnarray}
  \nu' &=& \nu e^\phi \\
  \sigma' &=& \sigma+\int_{|x|\le1}x(\nu'-\nu) (\d x)+a \eta
\end{eqnarray}

\end{proposition}
Un exemple, qui sera par la suite, à la base de la construction de la MME, est donné par la \emph{transformée d'Esscher}.

 \subsection{Transformée d'Esscher}\label{sec-transfE}
Une des méthodes populaires pour trouver une MME dans le cas des modèle de Lévy est d'utiliser la \emph{transformée d'Esscher}. Pour une distribution de probabilité $F(x)$, et pour $\theta$ un paramètre réel tel que $\int_\bbr e^{\theta y} \d F(y) <\infty$,  la transformée d'Esscher $F^\theta(x)$ est définie par :
\begin{equation}\label{form-Esscher-Transform}
\d F^\theta (x)=\frac{e^{\theta x}\d F(x)}{\int_\bbr e^{\theta y}\d F(y)}
\end{equation}
Si la distribution $F$ admet une densité alors $F^\theta$ admet également une densité
\begin{equation}\label{form-densite-Esscher}
  f^\theta (x)=\frac{e^{\theta x}f(x)}{\int_\bbr e^{\theta y} f(y) \d y}
\end{equation}
Pour plus d'informations sur la transformée d'Esscher on peut se référer à \cite{Ess32} et \cite{BE65}.

 \bigskip

 La transformée d'Esscher pour une mesure de probabilités, peut être définie de façon analogue. Étant donné un espace de probabilité $(\Omega,\cF,\P)$, et une variable aléatoire $X$, et un paramètre $\theta$; la transformée d'Esscher $\P^\theta$ (ou la mesure d'Esscher) est définie par
 \begin{equation}\label{form-Esscher-measure}
 \d \P^\theta=\frac{e^{\theta X} \d \P}{\E [e^{\theta X}]}
 \end{equation}
 A condition que l'espérance existe.

 \bigskip

Dans le cadre de modèles de Lévy, il existe une infinité de mesures martingales $Q$ telles que $X_t$ reste un processus de Lévy et satisfait
$$
e^{rt}=E_Q(e^{X_t})=exp \{t \kappa_Q(1)\}
$$
où $\kappa_Q$ est l'exposant caractéristique de
$X_t$ sous $Q$. $\kappa_Q$ est sous la forme
\begin{equation}\label{form-kappaQ}
    \kappa_Q(x)=\kappa(x+\theta)-\kappa(\theta)
\end{equation}

Lorsque l'équation $\kappa_Q(1)=r$ admet une seule solution $\theta^*$ l'exposant caractéristique $\kappa_Q$ est appelé \emph{la transformée d'Esscher}. L'équation peut ne pas admettre de solution (cas de queues de distribution épaisses).

\begin{theorem}
  On suppose $T>0$ et qu'il existe $\theta^*\in \bbr$ tel que
  $$
  \kappa(\theta^*+1)-\kappa(\theta^*)=0
  $$
\end{theorem}
avec $\E [e^{\theta^*X_T}]<\infty$ et $\E [e^{(\theta^*+1)X_T}]<\infty$, alors:
\begin{equation}\label{form-MME-P}
\frac{\d P^*}{\d P}=e^{\theta^*X_T-\kappa (\theta^*)T}
\end{equation}
définit une MME (mesure martingale équivalente) pour $(S_t)_{0\le t\le T}$. Le processus $X_t$ est un processus de Lévy sous $P^*$ et sont triplet caractéristique est $(a^*,\sigma^*,\nu^*)$ avec
\begin{eqnarray}
  a^* &=& a+\sigma \theta^*+\int (e^{\theta^*x}-1)x 1_{|x|\le 1}\nu(\d x) \\
  \sigma^* &=& \sigma \\
  \nu^*(\d x) &=& e^{\theta^*x}\nu(\d x)
\end{eqnarray}

\begin{proof}
  Voir \cite{KS02} théorème 4.1.
\end{proof}

\subsection{Le processus Normal Inverse Gaussien (NIG)}\label{sec-NIG}

Un processus de Lévy $NIG(\alpha,\beta,\mu ,\delta) $ est un processus stochastique tel que $X_t \sim NIG(\alpha,\beta,\mu t,\delta t) $.  Nous rappelons la densité  $NIG(\alpha,\beta,\mu,\delta)$,  $\alpha \ge 0, \, \delta \ge 0$, $|\beta|\ge \alpha$, $\mu \in \bbr$, qui est définie par la formule

\begin{equation}\label{form-NIG-df}
  f(x)=\frac{\alpha}{\delta}\frac{K_1 \left(\alpha \sqrt{\delta^2+(x-\mu)^2}\right)} {\sqrt{\delta^2+(x-\mu)^2}} \exp \left\{\delta \sqrt{\alpha^2-\beta^2}+\beta(x-\mu)\right\}
\end{equation}

où $K_1$ est la fonction de Bessel modifiée de troisième type.

\bigskip

La processus NIG est un processus de saut dont la mesure de Lévy admet comme densité
$$
\nu(x;\alpha,\beta,\delta)=\frac{\delta \alpha}{\pi |x|} e^{\beta x} K_1(\alpha |x|)
$$
et son exposant caractéristique est donné par
$$
\kappa(\xi)=\mu \xi+ \delta \left( \sqrt{\alpha^2-\beta^2}-\sqrt{\alpha^2-(\beta+\xi)^2}\right)
$$

\noindent A l'aide de \eqref{form-kappaQ} on peut montrer (voir \cite{AlbrecherPredota2004}) que sous la mesure martingale équivalente d'Esscher, $X_t$ est un processus $NIG(\alpha,\beta^*,\mu,\delta)$ où

\begin{equation}\label{form-betaE}
  \beta^*=\beta+\theta^*=\frac{-1}{2}+\sqrt{\frac{\alpha^2 (\mu-r)^2}{\delta^2+(\mu-r)^2}-\frac{(\mu-r)^2}{4\delta^2}}
\end{equation}
Cette dernière propriété et la formule (\ref{form-call-eur}) permettent de calculer par simulations Monté-Carlo, la valeur d'une option de type européenne.

\bigskip
Dans le cas d'un call européen, et dans le cadre du modèle NIG, on peut même obtenir une formule explicite de type Black-Scholes (voir \cite{AlbrecherPredota2004}).
\begin{eqnarray}\label{form-call-explicite}
  C_t &=& e^{-r \, (T-t)}\E^*\left( (S_T-K)_+ | \cF_t \right) \nonumber\\[3mm]
    &=& S_t \int_{\ln (K/S_t)}^\infty NIG_{(\alpha,\beta+\theta^*+1,(T-t)\delta,(T-t)\mu)}(x) \d x\nonumber \\[3mm]
    &-& e^{r \, (T-t)} \, K \int_{\ln (K/S_t)}^\infty NIG_{(\ln(K/S_t);\alpha,\beta+\theta^*,(T-t)\delta,(T-t)\mu)}(x)\d x
\end{eqnarray}
Cette formule nous sera très utile pour des fins de comparaison, afin de valider le programme de simulation.

\subsection{Cas du processus Variance Gamma}\label{sec-VG}

L'un des exemples les plus simples de processus de Lévy, avec une intensité infinie de sauts, est le processus \emph{gamma}. C'est un processus à accroissements indépendants et stationnaires tel que pour tout $t\ge0$, $X_t\sim \Gamma (\lambda t, \gamma)$. Nous rappelons la densité de la loi gamma $\Gamma (\lambda, \gamma)$
$$
f(x)=\frac{\gamma^\lambda x^{\lambda-1} e^{-\gamma x}}{\Gamma(\lambda)}, \mbox{ } x>0
$$
Le processus gamma admet comme fonction caractéristique la fonction
$$
\E(e^{i u X_t})=(1-i u /\gamma)^{-\lambda t}
$$et la densité de la mesure de Lévy associée à ce processus est donnée par
$$
\nu(x)=\frac{\lambda e^{-\gamma x}}{x} 1_{x>0}
$$

\noindent A partir du processus gamma, on peut construire un autre processus de saut appelé le processus VG  (\emph{variance gamma}) \cite{MadanKonikov02}, \cite{Lewis01}. Ce nouveau processus $VG(x_0, \lambda,\gamma,\beta,\sigma)$ est obtenu en changeant l'échelle de temps d'un mouvement brownien avec drift, par un processus gamma:
\begin{equation}\label{form-Ytx0}
  Y_t=x_0+\beta X_t +\sigma B_{X_t}
\end{equation}

C'est donc un mouvement brownien sur une échelle de temps aléatoire donnée par le processus gamma. Le processus variance gamma est également un processus de Lévy avec intensité infinie de sauts.

\bigskip

La densité d'une loi $VG(x_0, \lambda,\gamma,\beta,\sigma)$  de paramètres $x_0\in \bbr, \, \lambda>0, \gamma>0, \beta \in \bbr$ et $\sigma>0$ est donnée par
$$
f(x)=\sqrt{\frac{2}{\pi \sigma^2}}\frac{\gamma^\lambda \, e^{\beta(x-x_0)/\sigma^2} (|x-x_0|/\sigma)^{\lambda-1/2}}{\Gamma(\lambda)\,  \left(\sqrt{\beta^2/\sigma^2+2 \gamma}\right)^{\lambda-1/2}}
K_{\lambda-1/2} \left( \frac{|x-x_0|}{\sigma}\sqrt{\beta^2/\sigma^2+2\gamma}\right)
$$

\bigskip

Dans le cas du modèle exponentiel de Lévy de type $VG (x_0, \lambda,\gamma,\beta,\sigma=1)$ ie: $Y_t \sim VG (x_0 t, \lambda t,\gamma,\beta,1)$ pour tout $t\ge 0$,  Hubalek et Sgarra \cite{HubaalekSgarra} ont calculé la MME à l'aide de la transformée d'Esscher. Il s'avère que si $\beta^2+2 \gamma \le \frac{1}{4}$ la mesure martingale d'Esscher $P^*$  du processus $e^Y$ n'existe pas. Si $\beta^2+2 \gamma > \frac{1}{4}$  la mesure d'Esscher $P^*$ existe et le paramètre d'Esscher est donné par la formule explicite:
$$
\theta^*=-\beta+\frac{-1+\sqrt{1+\beta^2 \ve^2 -\ve +2 \gamma \ve^2}}{\ve} \mbox{ où } \ve=1-e^{x0/\lambda}, \,\, x_0>0
$$
Le processus $Y$ sous $P^*$ est un processus  $VG(x_0,\lambda,\gamma^*, \beta^*,1)$ avec
$$
  \left\{
    \begin{array}{cl}
    \gamma^* &= \gamma-\beta \theta^*-{\theta^*}^2/2 \\[3mm]
    \beta^* &= \beta+\theta^*
    \end{array}
  \right.
$$ Grace donc à cette propriété, et à l'aide de la formule (\ref{form-call-eur}), on peut calculer la valeur d'une option de type européenne à l'aide de simulations Monté-Carlo.

\bigskip

Noter qu'un process variance gamma peut être écrit comme la différence de deux processus gamma indépendants
\begin{equation}\label{form-VGdiff}
  Y_t=G^+(t)-G^-(t)
\end{equation}

$G^+$ est un processus $gamma(\lambda=\frac{(\mu^+)^2}{\nu^+}, \gamma=\frac{\mu^+}{\nu^+})$, c'est à dire de paramètre de moyenne $\mu^+$ et de paramètre de variance $\nu^+$, et $G^-$ est un processus $gamma(\frac{(\mu^-)^2}{\nu^-}, \frac{\mu^-}{\nu^-})$ tel que :

\begin{eqnarray}
  \mu^+ &=& \left(\sqrt{\beta^2+2\sigma^2/\nu} +\beta \right)/2\\
  \mu^- &=&  \left(\sqrt{\beta^2+2\sigma^2/\nu} -\beta \right)/2\\
  \nu^+ &=& (\mu^+)^2 \nu \\
  \nu^- &=& (\mu^-)^2 \nu
\end{eqnarray}
Cette propriété est très intéressante, et sera utilisée pour les simulations.

\subsection{Une autres mesures martingales équivalentes}\label{sec-autres-MME}
La transformée d'Esscher n'est pas la seule méthode pour obtenir une MME pour le processus VG ou pour les processus NIG. On considère le modèle exponentiel de Lévy $S=\{S_t=S_0e^{Y_t}, \,t\ge 0 \}$ où $Y$ est un processus $VG(x_0=0, \lambda=1/\nu,\gamma=\nu,\beta,\sigma)$ (ie: que le subordinateur de (\ref{form-Ytx0}) est un processus gamma de paramètre de moyenne $\mu=1$ et paramètre de variance $\nu$); dans ce cas la fonction caractéristique sera donné par :
\begin{equation}\label{form-VG-caract-funct}
  \E\left(i u Y_t\right)=\left(\frac{1}{1-iu \beta \nu+\frac{1}{2}\sigma^2 \nu u^2} \right)^{t/\nu}
\end{equation}
On suivant \cite{EberleinKeller95}, \cite{BarndroffNielsenShephard00} et \cite{MadanCarrChang98}, sous la mesure risque neutre, le cours $S_t$ adment comme trajectoires
\begin{equation}\label{form-VG-St}
  S_t=S_0 e^{(r+\omega) t+Y_t}
\end{equation}
où $r$ est le rendement instantané de l'actif sans risque. La constante $\omega$ définie par l'équation $e^{-\omega}=\E \left(e^{X_1}\right)$, assure que les prix actualisé soient une martingale sous la mesure associée à l'accumulateur compte numéraire. De l'équation (\ref{form-VG-caract-funct}) on peut avoir une form explicite de $\omega$:
\begin{equation}\label{form-VG-omega}
  \omega=\frac{\log(1- \beta \nu -\sigma^2 \nu/2)}{\nu}
\end{equation}

\bigskip
Dans le cas des modèles NIG, un procédé similaire permet d'obtenir les trajectoires sous la mesure risque-neutre, qui seront sous la forme (\ref{form-VG-St}) avec
\begin{equation}\label{form-VG-omegaRN}
  \omega=-\mu-\delta \sqrt{\alpha^2-\beta^2}+\delta \sqrt{(\alpha^2-(1+\beta)^2)}
\end{equation}
Noter que dans ces deux cas le processus NIG et VG doivent commencer en  zero, (ie: dans la formule (\ref{form-Ytx0}) on aura $x_0= 0$)   et l'instant initial également doit être initialisé à  $t_0 = 0$.

\bigskip
Ainsi donc, pour l'évaluation des options de type européenne, on utilise l'expression sous la mesure risque neutre du processus $S_t$ (ie: la formule (\ref{form-VG-St})), et on utilise les simulation Monté-Carlo pour trouver le prix grace à la formule (\ref{form-call-eur}).

\section{Résultats numériques}\label{sec-simulation}

\subsection{Modèle exponentiel de Lévy de type NIG}
On souhaite comparer la valeur d'une option asiatique donné par la mesure martingale d'Esscher et celle donnée par la formule \eqref{form-VG-St} et \eqref{form-VG-omegaRN}. Pour ce fait on considère les donnée de marché suivantes. $T=1/12$ et $T=2/12$, $S_0=36$, $K=34,35,36,37...$ et les paramètres du processus NIG sont tirés de l'article \cite{AlbrecherPredota2004}; à savoir : $\alpha=81.6$, $\beta=3.69$, $\delta=0.0103$, $\mu=-0.000123$. Pour valider notre algorithme de simulation, on évalue la valeur d'un call européen, dans le modèle NIG, à la fois avec la méthode Monté-Carlé, en simulant les trajectoires de ce processus, mais également à l'aide la formule explicite donnée par (\ref{form-call-explicite}). Par défaut d'avoir une machine puissante, le nombre de simulation effectué à chaque fois est de $n=10000$. Les résulats sont répértoriés dans le tableau ci dessous.

\bigskip
\begin{tabular}{|c|c|c|c|c|c|c|c|}
\hline
\hline
   $S_0$	& T 	  &	 r &	 K &	 $C_0$ &	 $C_0 (MC)$  &	 $A_0 (Esscher)$  & 	 $A_0$  \\
\hline\hline
 36	& 1/12	& 0.1&	 $34.0$ &	  2.2822 &	 $2.227\pm0.07$ &	 $2.111\pm0.052$  &	 $2.148\pm0.001$  \\
 	& 	&	 &	 $35.0$ &	    1.2918 &	 $1.263\pm0.072$ &	 $1.146\pm0.045$  &	 $1.156\pm0.001$  \\
 	& 	&	 &	 $36.0$ &	 0.7962 &	 $0.229\pm0.049$ &	 $0.121\pm0.026$  &	 $0.165\pm0.001$  \\
 	& 	&	 &	 $37.0$ &	 0.812 &	 $0.234\pm0.101$ &	 $0.123\pm0.07$  &	 $0.0$  \\
 \hline
 & 	&	0.05 &	 $34.0$ &	   2.1414 &	 $2.135\pm0.028$ &	 $2.066\pm0.019$  &	 $2.074\pm0.001$  \\
 	& 	&	 &	 $35.0$ &	  1.1463 &	 $1.151\pm0.029$ &	 $1.084\pm0.018$  &	 $1.078\pm0.001$  \\
 	& 	&	 &	 $36.0$ &	  0.4366 &	 $0.152\pm0.023$ &	 $0.082\pm0.013$  &	 $0.085\pm0.001$  \\
 	& 	&	 &	 $37.0$ &	   0.7425 &	 $0.104\pm0.032$ &	 $0.051\pm0.02$  &	 $0.0$  \\
 \hline
 & 2/12	&0.1&  $34.0$ &	   2.5620 &	 $2.468\pm0.117$ &	 $2.23\pm0.085$  &	 $2.294\pm0.002$  \\
 	& 	&	 &	 $35.0$ &	   1.5797 &	 $1.529\pm0.119$ &	 $1.293\pm0.074$  &	 $1.31\pm0.002$  \\
 	& 	&	 &	 $36.0$ &	   1.0477 &	 $0.469\pm0.082$ &	 $0.248\pm0.044$  &	 $0.326\pm0.002$  \\
 	& 	&	 &	 $37.0$ &	   1.9032&	 $0.464\pm0.164$ &	 $0.244\pm0.114$  &	 $0.0$  \\
 \hline
 & 	&0.05	 &	 $34.0$ &	  2.2822 &	 $2.273\pm0.042$ &	 $2.135\pm0.027$  &	 $2.148\pm0.002$  \\
 	& 	&	 &	 $35.0$ &	  1.2912 &	 $1.299\pm0.043$ &	 $1.165\pm0.027$  &	 $1.156\pm0.002$  \\
 	& 	&	 &	 $36.0$ &	  0.5617 &	 $0.3\pm0.035$ &	 $0.162\pm0.02$  &	 $0.167\pm0.002$  \\
 	& 	&	 &	 $37.0$ &	  0.7927 &	 $0.204\pm0.046$ &	 $0.1\pm0.029$  &	 $0.0$  \\
\hline
\end{tabular}

\bigskip
\bigskip
Les résultats obtenus dans la colonne $C_0$ (valeur du call par la formule explicite) et celle de la colonne $C_0(MC)$ (valeur du call par simmulations Monté-Carlo) permettent de valider l'algorithme, même si il y'a quelques différences notamment lorsque $K$ est très proche de $S_0$. Des différences que j'expliques pour le le nombre réduit de simulations effectuées (seulement 10 000), mais également par l'imprécision dans le calcul de la fonction de répartition d'une loi NIG (à cause de la fonction de Bessel). La fonction de répartition de la loi NIG est utilisée pour le calcul explicite de la valeur du Call (formule \eqref{form-call-explicite}).
\bigskip

Les colonnes $A_0(Esscher)$ et $A_0$ répertorient les valeurs d'une option asiatique, obtenues avec les deux méthode; la transformée d'Esscher et la méthode décrite en section \ref{sec-autres-MME}. Les données récoltées permettent d'affirmer que les deux procédés donnent de manière générale la même valeur de l'option. Les différences entre les deux colonnes sont négligeables! On fait quand-même remarquer que pour le calcul qui ne fait pas intervenir la transformée d'Esscher, le prix $A_0$ observé, s'annule pour $K\ge 37$, alors qu'il est juste proche de zéro pour $K=37$, avec la transformée d'Esscher. Ce fait est probablement due au faible nombre de trajectoires utilisée dans les simulations. En fait, même en augmentant le nombre de trajectoires, on pourrait s'attendre, avec la croissance du prix d'exercice $K$, qu'un des deux algorithmes donne une valeur nulle, avant que l'autre ne le rejoignent en augmentant d'avantage la valeur de $K$. Ceci est prévisible, vue que sous les deux mesures martingales, les trajectoires des deux processus sont différentes. En d'autres termes, à partir d'une certaine valeur de $K$ suffisamment grande, aucune trajectoires ne réalisera un payoff positif; d'où la nullité de la valeur de l'option.

\subsection{Modèle exponentiel de Lévy de type VG}

Pour les simulation du modèle VG, on utilise les donnée de marché tirées de l'article \cite{Lecuyer09}. On a donc $T=1$, $s=16$ (nombre d'instants d'observations), $r=0.1$, $\beta=-0.1436$ (drift du MB), $\sigma=0.12136$, $\mu=1.0$ ( paramètre de la moyenne du processus Gamma) et $\nu=0.3$ est le paramètre de variance du processus gamma. On a également $K=101.0$ et $S0=100.0$. On peut valider l'algorithme de simulation à l'aide de l'artilce \cite{Lecuyer09}. On effet pour les même donnée de marché on obtient la valeur d'une option asiatique $A_0=5.528$ avec un intervalle de confiance de $(5.203, 5.853 )$. Dans l'artile de L'Ecuyer (\cite{Lecuyer09}), la valeur de $A_0=5.725$.

\bigskip
Dans cette partie, on aimerait bien comparer la valeur d'une option asiatique donnée par la mesure d'Esscher (section \ref{sec-VG}), et celle donné dans la section \ref{sec-autres-MME}. Pour cette objectif on fait l'hypothèse (irréaliste) que $\sigma=1$ et $\nu=1$ afin que les deux démarches, décrites dans les sections précédentes, soient valables, et les calculs des deux approches soient cohérents. On pose également $x_0=0.00000001$, une valeur très proche de zéro, comme point initial du processus VG, car le raisonnement en section \ref{sec-VG} (avec la transformée d'Esscher), n'est valable que pour $x_0 \ne 0$. Les résultats obtenus sont répertoriés dans le tableau ci dessous :

\bigskip
\begin{tabular}{|c|c|c|c|c|c||c|c|}
\hline
\hline
	 r &	 K &	 $A_0 (Esscher) BGSS$ &	 $A_0-BGSS$  &	 $A_0 (Esscher)-DG$  & 	 $A_0-DG$  \\
   \hline
   0.1  & 95  & 26.095   & 27.337   & 24.184  & 24.879\\
    & 101  & 23.823  & 25.289  &  21.897 & 22.811\\
     & 105 & 22.460  &  24.057 &  20.518 & 21.568\\
    0.05  & 95  & 25.532  & 26.895  & 23.592  & 24.401\\
     &  101 & 23.270  &  24.855  & 21.310  & 22.340\\
   &  105 & 21.916  & 23.627    & 19.942& 21.101\\
\hline\hline
\end{tabular}

\bigskip

On utilise deux méthodes différentes pour simuler un processus VG. La première méthode est la méthode séquentielle (BGSS) (Brownian gamma sequential sampling), qui utilise un processus gamma comme un subordinateur d'un brownien avec drift. L'autre méthode est la méthode DG (Différence gamma sampling) basée sur le fait qu'un processus VG peut s'écrire comme la différence de deux processus gamma indépendants (voir section \ref{sec-VG}).

\bigskip

On voit bien d'après les observations du tableau, que la différence du prix donné par la mesure d'Esscher, et le prix donné par l'autre mesure martingale basée sur la modification des trends du processus VG, donnent en général les mêmes prix d'options.

\section{Conclusion}\label{sec-conclusion}
Dans ce rapport on s'est intéressé à l'évaluation d'une option de type européenne, sous le modèle exponentiel de Lévy. On a choisit comme exemple l'option asiatique, mais on aurait pu prendre n'importe quel option de type européenne, sans en altérer la démarche. Le modèle qu'on a choisit sont les modèle NIG et VG. On ne sait pas toujours trouver la mesure d'Esscher pour ces modèles dans le cas général. Par example pour le modèle VG on avait besoin d'un processus qui ne démarre pas en $x_0=0$ et on avait posé $\sigma=1$. Ils serait bien d'avoir la mesure d'Esscher dans le cas plus général. On s'est intéressé à deux méthodes de construction d'une MME. La première est basée sur la transformée d'Esscher, et l'autre consiste à apporter une correction "risque-neutre" sur la dynamique des trajectoires. Les résultats numériques obtenus, nous disent que les deux méthodes produisent en général les même prix. Il serait intéressant de confirmer les résultats de ce travail avec d'autres type d'options, en faisant un choix très varié des différents paramètres, et en simulant un nombre élevé de trajectoires.
\bibliographystyle{plain}

\end{document}